\documentclass[
    letterpaper,
    reprint,
    aps,
    pra,
    superscriptaddress,
    showkeys,
    amssymb,
    english,
]{revtex4-2}

\usepackage[useregional]{datetime2}
\usepackage{comment}
\usepackage{mathtools}
\usepackage{csquotes}
\usepackage[unicode]{hyperref}
\hypersetup{
	colorlinks = true,
	allcolors = blue,
	bookmarksnumbered = true,
	bookmarksopen = true,
}
\usepackage[all]{hypcap}
\usepackage{microtype}
\usepackage{cleveref}

\newcommand*{\eqendpoint}{.}
\newcommand*{\eqcomma}{,}

\newcommand*{\decsign}{.}
\newcommand*{\unitvec}[1]{\mathbf{\widehat{#1}}}

\newcommand*{\ebcm}{EBCM}
\newcommand*{\vecdio}[1]{\mathbf{#1}}

\newcommand*{\conj}[1]{#1^{*}}
\newcommand*{\dyadic}[1]{\overset{\leftrightarrow}{#1}}
\newcommand*{\angconj}[1]{\overline{#1}}
\newcommand*{\real}{\text{Re}}
\newcommand*{\imag}{\text{Im}}

\newcommand*{\vecnabla}{\vecdio{\nabla}}

\newcommand*{\polconj}[1]{\widetilde{#1}}

\newcommand{\citewithref}[1]{Ref.~\cite{#1}}
\newcommand{\citewithrefs}[1]{Refs.~\cite{#1}}
\newcommand{\eqrefdio}[1]{Eq.~\eqref{#1}}
\newcommand{\eqrefs}[1]{Eqs.~\labelcref{#1}}
\newcommand{\figref}[1]{Fig.~\ref{#1}}
\newcommand{\figrefs}[1]{Figs.~\labelcref{#1}}
\newcommand{\appendixref}[1]{Appendix~\ref{#1}}
\newcommand{\subsectionref}[1]{Subsection~\ref{#1}}
\newcommand{\sectionref}[1]{Section~\ref{#1}}

\begin{document}

\title{Perturbative Born theory for light scattering
by time-modulated scatterers}
\date{April 20, 2026}

\author{Dionysios Galanis}
\affiliation{Institute of Nanoscience and Nanotechnology,
\href{https://ror.org/038jp4m40}{NCSR “Demokritos”},
Patriarchou Gregoriou and Neapoleos Str., Ag. Paraskevi,
Athens GR-153 10, Greece}
\affiliation{Section of Condensed Matter Physics,
\href{https://ror.org/04gnjpq42}{National and Kapodistrian University of Athens},
University Campus, GR-157 84 Athens, Greece}
\author{Evangelos Almpanis}
\author{Nikolaos Papanikolaou}
\affiliation{Institute of Nanoscience and Nanotechnology,
\href{https://ror.org/038jp4m40}{NCSR “Demokritos”},
Patriarchou Gregoriou and Neapoleos Str., Ag. Paraskevi,
Athens GR-153 10, Greece}
\author{Nikolaos Stefanou}
\affiliation{Section of Condensed Matter Physics,
\href{https://ror.org/04gnjpq42}{National and Kapodistrian University of Athens},
University Campus, GR-157 84 Athens, Greece}

\begin{abstract}
We present a theoretical framework for electromagnetic scattering
by particles with a permittivity that is periodically varying in time,
based on a perturbative approach. Within this framework, we derive
explicit expressions for the scattering matrix of the dynamic system
in a first-order Born approximation, relating it directly
to the corresponding static problem. We show that inelastic scattering amplitudes
are governed by overlap integrals between static modes
at the input and output frequencies. Using this insight,
we analyze scattering from a time-modulated,  isotropic, dielectric sphere
and a high-permittivity dielectric cylinder, and demonstrate
how modal orthogonality can suppress inelastic channels,
while appropriate tuning of geometric parameters can significantly enhance them.
In particular, we show that cylindrical resonators support
strong inelastic scattering when resonance-to-resonance optical transitions,
induced by the temporal variation, involve a high-Q supercavity mode.
Comparison with full time-Floquet calculations confirms
that the first-order Born approximation remains quantitatively accurate
for modest modulation amplitudes and provides clear physical intuition
for frequency conversion and resonance-mediated scattering processes
in time-modulated photonic resonators.
\end{abstract}


\maketitle

\section{\label{intro}Introduction}

The ability to control light in both space and time has opened
new avenues in optics and photonics, enabling functionalities
that go beyond those of the respective static systems.
Although the study of time-modulated electromagnetic media dates back
to the 1950s and 1960s~\cite{morgenthalerVelocityModulationElectromagnetic1958,%
cassedyDispersionRelationsTimespace1963}, it has only recently gained
renewed attention thanks to advances
in modern experimental and nanofabrication techniques.
Temporal modulation in photonic systems enables a wide range of applications,
including nonreciprocal propagation of light without the need of
magnetic materials~\cite{panagiotidisOpticalTransitionsNonreciprocity2023a},
frequency conversion using linear materials~\cite{
karlFrequencyConversionTimeVariant2020a},
parametric amplification~\cite{valdez-garciaParametricResonancesPhotonic2024,%
koutserimpasParametricAmplificationBidirectional2018}, and the realization
of photonic time crystals (systems with wavevector gaps)~\cite{
asgariTheoryApplicationsPhotonic2024}.
Beyond these fundamental effects,
time-modulated photonic platforms have been proposed for more sophisticated applications,
ranging from the creation of synthetic dimensions and the exploration of
topological phases~\cite{yuanSyntheticDimensionPhotonics2018}
to photonic analogs of the Aharonov–Bohm effect~\cite{
mukherjeeExperimentalObservationAharonovBohm2018},
temporal Wood anomalies~\cite{galiffiWoodAnomaliesSurfaceWave2020},
extension of the Smith-Purcell effect into the fourth
dimension~\cite{koutserimpasExtendingSmithPurcellEffect2025}
and the optical equivalent of Fresnel drag~\cite{
huidobroFresnelDragSpace2019}.
An extensive list of possible applications of time-modulated electromagnetic media
can be found in \citewithrefs{calozSpacetimeMetamaterialsPart2020,%
taravatiSpaceTimeModulationPrinciples2020,%
galiffiPhotonicsTimevaryingMedia2022,pacheco-penaTimevaryingElectromagneticMedia2022,%
hayranUsingTimeVaryingSystems2023,ptitcynTutorialBasicsTimeVarying2023,%
boltassevaPhotonicTimeCrystals2024}.

From a fundamental perspective, complex electromagnetic structures
can be regarded as assemblies of elementary building blocks,
each corresponding to a single scatterer. Such time-modulated scatterers
constitute the essential ingredients of dynamically reconfigurable photonic systems,
and their study provides valuable physical insight
that remains only partially explored. Typical examples
include particles with a time-varying permittivity~\cite{
stefanouLightScatteringSpherical2021,schabScatteringPropertiesSpherical2022,%
asadchyParametricMieResonances2022a,stefanouLightScatteringPeriodically2023a}
or geometrical deformation~\cite{panagiotidisInelasticLightScattering2022a},
where the temporal variation may arise
from electrical control~\cite{sorefElectroopticalEffectsSilicon1987},
acoustic vibrations~\cite{lombardiPulsedMolecularOptomechanics2018},
spin-wave excitations~\cite{almpanisDielectricMagneticMicroparticles2018,%
krichevskyInverseFaradayEffect2024}, or even
in an all-optical manner using the nonlinear Kerr effect
\cite{guoNonreciprocalMetasurfaceSpace2019a}.
Transparent conductive oxides also
offer a promising avenue for all-optical time-variation using
ultrashort laser pulses~\cite{alamLargeOpticalNonlinearity2016,%
sahaPhotonicTimeCrystals2023}. The temporal modulation of a scatterer
couples different frequency components of the electromagnetic field,
leading to the generation of frequency sidebands while the energy of the 
optical system is not conserved, since energy exchange with the 
external modulation source is allowed.
Depending on the modulation strength and frequency,
rich physical behavior can emerge, including asymmetric scattering spectra,
enhanced forward or backward scattering, and effective gain or loss
associated with the modulation process.

In order to provide a versatile and computationally
efficient description of time-modulated optical scatterers,
we develop here a perturbative Born theory in which the temporal variation
of the material permittivity is treated as a small parameter,
as is relevant for realistic applications. This approach
enables a systematic evaluation of the coupling between frequency components
induced by the variation, while retaining the full spatial dependence
of the scattering problem. Within this formalism,
the electromagnetic fields are expanded in harmonics
of the modulation frequency, and the first-order correction terms
naturally describe processes such as frequency conversion
and modulation-induced asymmetry in the scattering pattern.
Unlike previous, more accurate, descriptions using
time-Floquet expansions~\cite{stefanouLightScatteringSpherical2021,%
stefanouLightScatteringPeriodically2023a}
or fully numerical finite-difference
time-domain~(FDTD)~\cite{yeeNumericalSolutionInitial1966} simulations,
the perturbative treatment provides direct physical insight
into the mechanisms governing optical transitions between modes
and yields closed-form expressions
for the sideband amplitudes. Moreover, it can be readily incorporated
into existing multiple-scattering or modal-expansion
frameworks~\cite{panagiotidisOpticalTransitionsNonreciprocity2023a,%
wangExpandingMomentumBandgaps2025a,gargPhotonicTimeCrystals2025},
thereby bridging detailed models of individual dynamic scatterers
with macroscopic time-modulated photonic structures.

The remainder of the paper is organized as follows.
\sectionref{theory} presents a theoretical framework for modeling scattering
from bodies with periodically varying permittivity,
based on a perturbative approach and, for practical implementations,
restricted to the first-order Born approximation.
\sectionref{results} compares this approximation with the dynamic
Extended Boundary Condition Method (\ebcm)~\cite{stefanouLightScatteringPeriodically2023a}
and demonstrates that, at least for small modulation strengths,
inelastic scattering amplitudes are governed by the overlap
between static modes at the input and output frequencies.
Both a dielectric sphere and a high-permittivity dielectric cylinder
are treated as examples of inelastic scattering behavior
with a particular focus on resonance-to-resonance optical transitions.
\sectionref{conclusions} summarizes the main conclusions of the work.

\section{\label{theory}Theoretical Methods}
In this section, we present a theoretical method
for light scattering by a single particle with a permittivity
that varies periodically in time, employing a perturbative approach
and deriving the scattering $T$~matrix within the first-order Born approximation.
For completeness and comparison, we first provide a 
brief review of the full time-Floquet method developed previously
for spherical scatterers~\cite{stefanouLightScatteringSpherical2021,%
panagiotidisInelasticLightScattering2022a} and 
scatterers of general shape~\cite{stefanouLightScatteringPeriodically2023a}.  

\subsection{\label{subsection:ebcm}Time-Floquet method}
We consider a finite scatterer occupying volume $V_{\text{in}}$,
made of a homogeneous and isotropic material with permeability
$\mu\mu_{0}$ and periodically varying permittivity $\varepsilon(t)\varepsilon_{0}$ surrounded
by a homogeneous, isotropic and time-independent medium with permeability
$\mu_{\text{h}}\mu_{0}$ and permittivity $\varepsilon_{\text{h}}\varepsilon_{0}$.
Due to Floquet's theorem~\cite{floquetEquationsDifferentiellesLineaires1883},
any solution of Maxwell's equations 
can be written as a superposition of Floquet solutions of the form
\begin{equation}
\vecdio{E}(\vecdio{r}, t)
= \mathrm{e}^{-i\omega t}\sum_{n} \vecdio{E}_{n}(\vecdio{r}) \mathrm{e}^{in\Omega t} \eqcomma
\label{eq:total_fields}
\end{equation}
where $\omega$ is the Floquet quasifrequency, $\Omega$ is the modulation frequency,
and the sum extends over all integers. The magnetic field can be expanded similarly.
Then, we expand each divergence-less 
$\vecdio{E}_{n}(\vecdio{r})$ in transverse vector spherical waves.

Away from the scatterer, the field is expanded as
\begin{equation}
\vecdio{E}^{\text{out}}_{n}(\vecdio{r})
= \vecdio{E}^{\text{inc}}_{n}(\vecdio{r})
+ \vecdio{E}^{\text{sc}}_{n}(\vecdio{r})\eqcomma
\label{eq:external_field}
\end{equation}
which is the sum of an incoming field, typically a plane wave,
\begin{equation}
\vecdio{E}^{\text{inc}}_{n}(\vecdio{r}) = \sum_{L} a_{Ln}^{\text{inc}} \;
\vecdio{J}_{L}(q_{\text{h}n}, \vecdio{r})\eqcomma
\label{eq:incoming_field}
\end{equation}
and the scattered field
\begin{equation}
\vecdio{E}^{\text{sc}}_{n}(\vecdio{r})
= \sum_{L} a_{Ln}^{\text{sc}} \;
\vecdio{H}_{L}(q_{\text{h}n}, \vecdio{r})\eqcomma
\label{eq:scattered_field}
\end{equation}
where
\begin{equation}
q_{\text{h}n} = \frac{\sqrt{\mu_{\text{h}}\varepsilon_{\text{h}}}}{c}(\omega - n\Omega)
= \sqrt{\mu_{\text{h}}\varepsilon_{\text{h}}}k_{n}\eqcomma
\end{equation}
$c$ being the velocity of light in vacuum,
while $L$ is a shorthand index for the indices $\ell, m$ and 
polarization modes $P = E, H$.
As in \citewithref{stefanouLightScatteringPeriodically2023a},
$\vecdio{J}_{H\ell m}(q_{\text{h}n}, \vecdio{r})
= j_{\ell}(q_{\text{h}n}r) \vecdio{X}_{\ell m}(\unitvec{r})$ and
$\vecdio{J}_{E\ell m}(q_{\text{h}n}, \vecdio{r})
= \frac{i}{q_{\text{h}n}} \vecnabla \times
\vecdio{J}_{H\ell m}(q_{\text{h}n}, \vecdio{r})$
are the regular transverse vector spherical waves, $j_{\ell}(x)$ are
the spherical Bessel functions, $\vecdio{X}_{\ell m}(\unitvec{r})$
are the vector spherical harmonics and 
$\vecdio{H}_{L}(q_{\text{h}n}, \vecdio{r})$ are the
outgoing transverse vector spherical waves, obtained by replacing
$j_{\ell}(q_{\text{h}n}r)$ in $\vecdio{J}_{L}(q_{\text{h}n}, \vecdio{r})$
with the spherical Hankel function of the first kind $h^{+}_{\ell}(q_{\text{h}n}r)$.
We emphasize that $\mathbf{H}_L$ should not be confused
with the magnetic field $\mathbf{H}$, as its meaning is clear from the context.
Inside the scatterer, the field $\vecdio{E}^{\text{in}}_{n}$
can also be expanded in transverse vector spherical
waves, as detailed in \citewithref{stefanouLightScatteringSpherical2021}.

Due to the linearity of Maxwell's equations, the coefficients $a_{Ln}^{\text{inc}}$
and $a_{Ln}^{\text{sc}}$ are linearly dependent,
and we define the $T$~matrix as
\begin{equation}
a_{Ln}^{\text{sc}} = \sum_{L', n'}T_{Ln; L'n'} a_{L'n'}^{\text{inc}}\eqendpoint
\label{eq:tmatrix_definition}
\end{equation}
The $T$~matrix can be efficiently calculated for finite scatterers
close to a spherical shape using the dynamic \ebcm\ by
invoking the continuity of the components of the electric and magnetic
fields parallel to the scatterer's surface
\begin{subequations}
\label{eq:boundary_conditions}
\begin{align}
\unitvec{n} \times \vecdio{E}^{\text{in}}_{n} &= \unitvec{n}
\times \vecdio{E}^{\text{out}}_{n}\eqcomma \\
\unitvec{n} \times \vecdio{H}^{\text{in}}_{n} &= \unitvec{n}
\times \vecdio{H}^{\text{out}}_{n}\eqendpoint
\end{align}
\end{subequations}
Details of the derivation are given in
\citewithref{stefanouLightScatteringPeriodically2023a}.
In the end, the $T$~matrix is given by a system of linear equations
\begin{equation}
\sum_{L''n''} T_{Ln; L''n''}Q^{0}_{L''n''; L'\nu} = - Q_{Ln; L'\nu}^{+}\eqendpoint
\label{eq:tmatrix_solution}
\end{equation}
Explicit expressions for the matrix elements $Q_{Ln; L'\nu}^{0/+}$ are 
given in \citewithref{stefanouLightScatteringPeriodically2023a}.
This system of linear equations can be solved numerically
by introducing cutoffs in indices $\ell$ and $n$. The static
case is retrieved by imposing a cutoff in the Floquet components
equal to $n=0$. Solving for the transposed 
matrices improves numerical stability~\cite{mishchenkoScatteringAbsorptionEmission2006}.

\subsection{\label{subsection:born_general}Perturbative approach}
We now proceed with a perturbative approach to evaluate the $T$~matrix.
We consider the same finite scatterer as in \subsectionref{subsection:ebcm}
with periodically varying permittivity.
From Maxwell's equations in the absence of sources,
\begin{subequations}
\begin{align}
    \vecnabla \times \vecdio{E}(\vecdio{r}, t)
    &= -\frac{\partial}{\partial t} \vecdio{B}(\vecdio{r}, t)\eqcomma\\
    \vecnabla \times \vecdio{H}(\vecdio{r}, t)
    &= \frac{\partial}{\partial t} \vecdio{D}(\vecdio{r}, t)\eqcomma
\end{align}
\end{subequations}
it can be shown that the electric field both inside and outside the scatterer
satisfies
\begin{equation}
\label{eq:maxwell_electric}
    \vecnabla \times \frac{1}{\mu(\vecdio{r})} \vecnabla
    \times \vecdio{E}_{n}(\vecdio{r})
    = \left(\frac{\omega - n\Omega}{c}\right)^{2}
    \sum_{n'}\varepsilon_{n - n'}(\vecdio{r})\vecdio{E}_{n'}(\vecdio{r})\eqcomma
\end{equation}
where $\varepsilon$ and $\mu$ are allowed to vary spatially, and the Fourier transform
\begin{subequations}
\begin{align}
\varepsilon(\vecdio{r}, t) &= \sum_{n}\varepsilon_{n}(\vecdio{r}) \mathrm{e}^{in\Omega t}\eqcomma\\
\varepsilon_{n}(\vecdio{r}) &= \frac{1}{T}\int_{0}^{T} 
\varepsilon(\vecdio{r}, t) \mathrm{e}^{-in\Omega t} \, \mathrm{d}t
\end{align}
\end{subequations}
has been taken.

The first-order Born approximation is extracted from \eqrefdio{eq:maxwell_electric}
by treating the static case as the unperturbed system and the variation
as the perturbation. To be more specific, writing
\begin{equation}
    \varepsilon_{n}(\vecdio{r}) = \varepsilon^{(0)}(\vecdio{r})\delta_{n0}
    + \Delta\varepsilon_{n}(\vecdio{r})\eqcomma
\end{equation}
we have
\begin{equation}
\label{eq:maxwell_electric_perturbation}
\begin{split}
    &\vecnabla \times \frac{1}{\mu(\vecdio{r})} \vecnabla
    \times \vecdio{E}_{n}(\vecdio{r})
    - k_{n}^{2} \varepsilon^{(0)}(\vecdio{r}) \vecdio{E}_{n}(\vecdio{r})\\
    &= k_{n}^{2} \sum_{n'} \Delta\varepsilon_{n - n'}(\vecdio{r})
    \vecdio{E}_{n'}(\vecdio{r})\eqendpoint
\end{split}
\end{equation}
We now introduce the dyadic Green's function
$\dyadic{G}_{0}(k, \vecdio{r}, \vecdio{r}')$ for the static scatterer,
\begin{equation}
    \left(k^{2} \varepsilon^{(0)}(\vecdio{r})
    - \vecnabla \times \frac{1}{\mu(\vecdio{r})} \vecnabla
    \times \right)\dyadic{G}_{0}(k, \vecdio{r}, \vecdio{r}')
    = \dyadic{I}\delta(\vecdio{r} - \vecdio{r}')\eqendpoint
\end{equation}
This function can be shown~\cite{chewWavesFieldsInhomogenous1999,%
gonisGreenFunctionsOrdered1992} to be equal to
\begin{equation}
\label{eq:green_dyadic_sphere}
    \dyadic{G}_{0}(k, \vecdio{r}, \vecdio{r}')
    = -iq_{\text{h}}\mu_{\text{h}}\sum_{L}\vecdio{I}_{L}(k, \vecdio{r})
    \otimes \angconj{\vecdio{R}}_{L}(k, \vecdio{r}'),\quad r > r'\eqcomma
\end{equation}
where $\vecdio{I}_{L}(k, \vecdio{r})$ is the irregular solution,
equal to an outgoing wave $\vecdio{H}_{L}(q_{\text{h}}, \vecdio{r})$
outside the sphere, while $\angconj{\vecdio{R}}_{L}(k, \vecdio{r})$
is the regular solution, obtained by the scattering of an incoming
wave $\angconj{\vecdio{J}}_{L}(q_{\text{h}}, \vecdio{r})$.
The \emph{bar} operation on a vector spherical wave is defined as follows~\cite{stefanouLightScatteringPeriodically2023a}
\begin{subequations}
	\label{eq:angular_conjugation}
	\begin{align}
		 \angconj{\vecdio{F}}_{H\ell m} &= f_{\ell}\conj{\vecdio{X}_{\ell m}}
         = (-1)^{m + 1} \vecdio{F}_{H\ell -m}\eqcomma \\
		 \angconj{\vecdio{F}}_{E\ell m} &= -\frac{i}{q} \vecdio{\nabla}
		\times \left(f_{\ell}\conj{\vecdio{X}_{\ell m}}\right)
        = (-1)^{m} \vecdio{F}_{E\ell -m}\eqcomma
	\end{align}
\end{subequations}
where $F$ can refer to Bessel or Hankel vector spherical waves, while $f$,
correspondingly, refer to the spherical Bessel or Hankel functions.
Moreover, if we introduce an integer index $p$ such that
\begin{equation}
	P = E \Leftrightarrow p = 0, \qquad P = H \Leftrightarrow p = 1\eqcomma
\end{equation}
and the \emph{bar} operation on index $L$ such that
\begin{equation}
	L = P\ell m \Leftrightarrow \angconj{L} = P, \ell, -m\eqcomma
\end{equation}
we can write the definition more compactly as
\begin{equation}
	\angconj{\vecdio{F}}_{L} = (-1)^{m + p} \vecdio{F}_{\angconj{L}}\eqendpoint
	\label{eq:angconj_def}
\end{equation}

By virtue of the Green's function and
\eqrefdio{eq:maxwell_electric_perturbation},
we obtain a Lippmann-Schwinger equation,
\begin{equation}
\label{eq:electric_lippmann}
\begin{split}
    &\vecdio{E}_{n}(\vecdio{r}) = \vecdio{E}^{(0)}_{n}(\vecdio{r})\\
    &- \int k_{n}^{2} \dyadic{G}_{0}(k_{n}, \vecdio{r}, \vecdio{r}')
    \sum_{n'} \Delta\varepsilon_{n - n'}(\vecdio{r}')
    \vecdio{E}_{n'}(\vecdio{r}') \, \mathrm{d}^{3}r' \eqcomma
\end{split}
\end{equation}
with $\vecdio{E}^{(0)}_{n}$ being a scattering
solution of the static system.

To calculate the $T$~matrix, we consider an incoming
vector spherical wave $\vecdio{J}_{L}(q_{\text{h}n}, \vecdio{r})$
of frequency $\omega - n\Omega$, so that in the perturbed system we have
\begin{equation}
    \label{eq:electric_field_tmatrix}
	\begin{split} \vecdio{E}_{n'}(\vecdio{r}) =
		&\vecdio{J}_{L}
        (q_{\text{h}n}, \vecdio{r})\delta_{nn'}\\
		&+ \sum_{L'}T_{L'n', Ln}
        \vecdio{H}_{L'}(q_{\text{h}n'},
        \vecdio{r}),\quad r>r_{>}\eqcomma\end{split} 
\end{equation}
$r_{>}$ being the radius of the smallest circumscribing sphere
of the scatterer, centered at the origin of our coordinate
system~\cite{mishchenkoScatteringAbsorptionEmission2006}.
On the other hand, the corresponding solution,
$\vecdio{E}_{n}^{(0)}$, for the unperturbed system is the regular solution
$\vecdio{R}_{L}(k_{n}, \vecdio{r})$ as in \eqrefdio{eq:green_dyadic_sphere}
(but without the bar)
\begin{equation}
\label{eq:electric_field_static_gen}
\begin{split}
    \vecdio{E}_{n}^{(0)}(\vecdio{r})
    &= \vecdio{R}_{L}(k_{n}, \vecdio{r}) \\ &= \begin{dcases}
		\begin{aligned}&\vecdio{J}_{L}(q_{\text{hn}}, \vecdio{r})\\
		&+ \sum_{L'}T_{L'L}^{(0)}(k_{n})
        \vecdio{H}_{L'}(q_{\text{hn}}, \vecdio{r})\end{aligned},
		\  r > r_{>} \\
		\sum_{L'}\alpha_{L'L}(k_{n}) \vecdio{J}_{L'}(q_{n}, \vecdio{r}),
		\quad \vecdio{r} \in V_{\text{in}}
	\end{dcases}\eqcomma
\end{split}
\end{equation}
with $\vecdio{E}_{n'}^{(0)} = 0$ for $n' \neq n$, and $q_{n} = k_{n}\sqrt{\mu\varepsilon}$.
Substituting $\angconj{L}$ for $L$ in \eqrefdio{eq:electric_field_static_gen},
using \eqrefdio{eq:angconj_def} and invoking the uniqueness of solutions
of Maxwell's equations, we obtain
\begin{equation}
\label{eq:regular_sol_gen}
\begin{split}
    \angconj{\vecdio{R}}_{L}(k_{n}, \vecdio{r})
    &= (-1)^{m + p}\vecdio{R}_{\angconj{L}}(k_{n}, \vecdio{r})\\
    &= \begin{dcases}
		\begin{aligned}&\angconj{\vecdio{J}}_{L}(q_{\text{hn}}, \vecdio{r})\\
		&+ \sum_{L'}\angconj{T}_{L'L}^{(0)}(k_{n})
        \vecdio{H}_{L'}(q_{\text{hn}}, \vecdio{r})\end{aligned},
		\  r >r_{>}\\
		\sum_{L'}\angconj{\alpha}_{L'L}(k_{n}) \vecdio{J}_{L'}(q_{n}, \vecdio{r}),
		\quad \vecdio{r} \in V_{\text{in}}
	\end{dcases}\eqcomma
\end{split}
\end{equation}
where
\begin{subequations}
\begin{align}
    \angconj{T}^{(0)}_{L'L}(k_{n})
    &= (-1)^{m + p} T^{(0)}_{L'\angconj{L}}(k_{n})\eqcomma\\
	\angconj{\alpha}_{L'L}(k_{n})
    &= (-1)^{m + p} \alpha_{L'\angconj{L}}(k_{n})\eqendpoint
	\label{eq:ainsbardef}
\end{align}
\end{subequations}
The first-order Born approximation
consists in taking $\vecdio{E}_{n}
= \vecdio{E}_{n}^{(0)}$ on the right
hand side of \eqrefdio{eq:electric_lippmann}.
Taking into account
\eqrefs{eq:green_dyadic_sphere,eq:electric_field_static_gen,eq:regular_sol_gen},
we get that, for $r$ large enough,
\begin{equation}
\label{eq:born_almostthere}
\begin{split}
    \vecdio{E}_{n'}(\vecdio{r}) = &\left\{\vecdio{J}_{L}
    (q_{\text{h}n}, \vecdio{r}) + \sum_{L'}T_{L'L}^{(0)}(k_{n})
    \vecdio{H}_{L'}(q_{\text{hn}}, \vecdio{r})\right\}\delta_{nn'}\\
    &+ \sum_{L'}
    \vecdio{H}_{L'}(q_{hn'}, \vecdio{r})
    \Big\{iq_{hn'}\mu_{h}k_{n'}^{2}\\
    &\times \sum_{L'', L'''} \Big[\angconj{\alpha}_{L''L'}(k_{n'})
    \alpha_{L'''L}(k_{n})\\&\times\int_{V_{\text{in}}}  \Delta\varepsilon_{n' - n}(\vecdio{r}')
    \vecdio{J}_{L''}(q_{n'}, \vecdio{r}') \cdot
    \vecdio{J}_{L'''}(q_{n}, \vecdio{r}') \, \mathrm{d}^{3}r' \Big]\Big\}\eqendpoint
\end{split}
\end{equation}
Comparing \eqrefdio{eq:born_almostthere} with
\eqrefdio{eq:electric_field_tmatrix}, we finally obtain
the first-order Born approximation for the $T$~matrix
(note the exchange of primed and unprimed indices)
\begin{equation}
\label{eq:born_tmatrix_nodisp}
\begin{split}
	T^{(1)}_{Ln; L'n'} = &i q_{\text{h}n}^{3}
    \int_{V_{\text{in}}}
    \frac{\Delta\varepsilon_{n - n'}(\vecdio{r})}{\varepsilon_{\text{h}}}
    \angconj{\vecdio{R}}_{L}(k_{n}, \vecdio{r}) \cdot \vecdio{R}_{L'}(k_{n'}, \vecdio{r}) \, \mathrm{d}^{3}r\\
    = &i q_{\text{h}n}^{3}
    \sum_{L'', L'''}\Big\{
		\angconj{\alpha}_{L''L}(k_{n})\alpha_{L'''L'}(k_{n'})\\
		&\times\int_{V_{\text{in}}} \frac{\Delta\varepsilon_{n - n'}(\vecdio{r})}{\varepsilon_{\text{h}}}
        \vecdio{J}_{L''}(q_{n}, \vecdio{r})
        \cdot \vecdio{J}_{L'''}(q_{n'}, \vecdio{r}) \, \mathrm{d}^{3}r \Big\}\eqendpoint
\end{split}
\end{equation}
In the case of spatially homogeneous modulation inside the scatterer,
$\Delta\varepsilon_{n - n'}$ can be extracted from the integral.
However, keeping the integral in its general form is advantageous,
as the perturbative approach—despite being approximate—allows one
to treat spatially inhomogeneous modulations,
contrary to the exact time-Floquet method outlined in \subsectionref{subsection:ebcm}.

It should be noted that the $T$~matrix
given by~\eqrefdio{eq:born_tmatrix_nodisp} is defined
by the solutions of the static particle corresponding
to incoming and outgoing waves with wavenumbers
$q_{\text{h}n}$ and $q_{\text{h}n'}$, respectively.
In general, the coefficients $\alpha_{L'''L'}(k_{n'})$
can be found by the static \ebcm. 
By virtue of \eqrefdio{eq:ainsbardef}, no extra effort is required
to calculate $\angconj{\alpha}_{L''L}(k_{n})$.
To reduce computational cost, the volume integral
can be converted into a surface integral, as shown
in \appendixref{appendix:overlap}.
An alternative way to derive the Born approximation for general Floquet media
can be obtained by generalizing the analysis of \citewithref{bothResonantStatesTheir2022}.

\subsection{\label{subsection:born_sphere}Application for a sphere}
In the case of a spherical scatterer of radius $R$,
a full analytical result can be obtained.
Since the expansion coefficients inside the sphere
have the form $\alpha_{LL'} = \alpha_{P\ell}\delta_{LL'}$, we have
\begin{subequations}
\label{eq:regular_sol_sphere}
\begin{align}
    \vecdio{R}_{L}(k, \vecdio{r})
    &= \alpha_{P\ell}(k) \vecdio{J}_{L}(q, \vecdio{r})\eqcomma \label{eq:sphereR}\\
    \angconj{\vecdio{R}}_{L}(k, \vecdio{r})
    &= \alpha_{P\ell}(k) \angconj{\vecdio{J}}_{L}(q, \vecdio{r}) \label{eq:sphereRbar}\eqcomma
\end{align}
\end{subequations}
for $r < R$. Substituting $\angconj{L}$ for $L$
in \eqrefdio{eq:sphereR}, by virtue of \eqrefs{eq:angconj_def,eq:regular_sol_gen},
we get \eqrefdio{eq:sphereRbar}.
The expressions for the expansion coefficients inside the sphere,
$\alpha_{P\ell}(k)$, are given in \appendixref{appendix:sphere}.
The overlap integral can also be given in a closed form,
as shown in \appendixref{appendix:sphere}.
Therefore, for a spatially homogeneous modulation, the $T$~matrix simplifies to
\begin{equation}
\label{eq:born_tmatrix_sphere_nodisp}
\begin{split}
	T^{(1)}_{Ln; L'n'} = &i q_{\text{h}n}^{3} \frac{\Delta\varepsilon_{n - n'}}{\varepsilon_{\text{h}}}
    \int_{V_{\text{in}}} \angconj{\vecdio{R}}_{L}(k_{n}, \vecdio{r}) \cdot 
    \vecdio{R}_{L'}(k_{n'}, \vecdio{r}) \, \mathrm{d}^{3}r\\
    = &i (q_{\text{h}n}R)^{3}
    \frac{\Delta\varepsilon_{n - n'}}{\varepsilon_{\text{h}}}\delta_{LL'}\\
	&\times \alpha_{P\ell}(k_{n})\alpha_{P\ell}(k_{n'}) D_{P\ell}(q_{n}R, q_{n'}R)\eqcomma
\end{split}
\end{equation}
where the radial overlap integrals $D_{P\ell}$ are given in
\appendixref{appendix:sphere}. 
The $\delta_{L'L}$ term indicates that modes with different $L=P \ell m$
do not couple, which is a selection rule, in the case
of time-modulated isotropic spheres, for the permittivity-variation
induced Mie-mode-to-Mie mode coupling. The same selection rule results,
also, from group-theoretical arguments. 
Since the symmetry of the Mie modes of the spherical particle
is that of the $O(3)$ Lie group, and the temporal perturbation
is an irreducible tensor operator which has the symmetry
of the $D_{\text{g}}^{(\ell=0)}$ irreducible representation of $O(3)$,
then the perturbation operating on an eigenvector
of the $P\ell$ irreducible subspace, transforms according to
the relevant direct product representation:
$D_{\text{g}}^{(\ell=0)}\otimes D_{\text{g,u}}^{(\ell)} = D_{\text{g,u}}^{(\ell)}$.
This implies a straightforward selection rule, according to
which Mie-mode-to-Mie mode photonic transitions conserve
both the angular momentum $\ell$ and the parity $\sigma=\text{g},\text{u}$ or,
equivalently, the polarization of the mode.

For particles of revolution that possess a mirror plane
perpendicular to the symmetry axis, such as spheroids or cylinders,
the continuous rotational symmetry is reduced from $O(3)$ to $D_{\infty h}$,
and the spherical Mie modes are split
accordingly~\cite{almpanisNonsphericalOptomagnonicResonators2021}.
In this case, the eigenmodes are classified according to
the irreducible representations of the $D_{\infty h}$ group,
labeled by $|m|$ and parity $\sigma=\text{g},\text{u}$. As for the sphere,
successive solutions within each irreducible subspace are indexed
by the radial order. Mode-to-mode transitions are allowed only between eigenmodes
belonging to the same irreducible subspace but with different radial orders.

\subsection{\label{subsection:cs}Cross sections}
In order to derive observable quantities from the $T$~matrix,
such as absorbed or scattered power, we need to calculate
the Poynting vector $\vecdio{S}$. In particular, we consider
the temporal mean value $\frac{1}{T} \int_{0}^{T} \, \mathrm{d}t$
over a very long time interval $T \rightarrow \infty$,
and then integrate over the surface of a sphere
$S^{2}(r)$, $r \rightarrow \infty$, that surrounds the scatterer
to obtain outgoing power. From there,
one can define normalized scattering
and absorption cross sections~\cite{stefanouLightScatteringPeriodically2023a}.
For an incident plane wave of unit amplitude
\begin{equation}
\label{eq:plane_wave}
\begin{split}
\vecdio{E}^{\text{inc}}(\vecdio{r}, t) &= \mathrm{e}^{-i(\omega - p\Omega)t}
\mathrm{e}^{iq_{\text{h}p}\unitvec{k}
\cdot \vecdio{r}}\unitvec{p}\\
&= \mathrm{e}^{-i(\omega - p\Omega)t}\sum_{L}a_{Lp}^{\text{inc}}
\vecdio{J}_{L}(q_{\text{h}p}, \vecdio{r})\eqcomma
\end{split}
\end{equation}
we get the (normalized) scattering cross section,
\begin{equation}
\sigma^{\text{sc}} = \sum_{n} \sigma^{\text{sc}}_{n}
= \frac{1}{\pi}\sum_{n, L}\frac{1}{(q_{\text{h}n}R_{\text{eff}})^{2}}
|a_{Ln}^{\text{sc}}|^{2}\eqcomma
\end{equation}
and the absorption cross section,
\begin{equation}
\sigma^{\text{abs}} = \sigma^{\text{ext}} - \sigma^{\text{sc}}\eqcomma
\end{equation}
where the extinction cross section is defined as
\begin{equation}
\sigma^{\text{ext}} = -\frac{1}{\pi}\sum_{L}\frac{1}{(q_{\text{h}p}R_{\text{eff}})^{2}}
\real \left[\conj{(a_{Lp}^{\text{inc}})}a_{Lp}^{\text{sc}}\right]\eqendpoint
\end{equation}
The effective radius $R_{\text{eff}}$ is an arbitrary normalization quantity.
In \citewithref{stefanouLightScatteringPeriodically2023a},
it is defined as the radius of a sphere with the same volume as the scatterer.
Here, we adopt a different convention: for spherical scatterers,
$R_{\text{eff}}$ is taken to be the sphere radius,
while for cylindrical scatterers, it is taken to be the cylinder radius.
In the case of spheres, the scattering cross section
takes a simpler form~\cite{stefanouLightScatteringSpherical2021}
\begin{equation}
    \sigma^{\text{sc}} = \sum_{n} \sum_{Pl} \sigma^{\text{sc}}_{P\ell n}
    = \sum_{n} \frac{2}{(q_{\text{h}n}R_{\text{eff}})^{2}}
    \sum_{Pl} (2l + 1)|T_{P\ell; np}|^{2}\eqcomma
    \label{eq:scs_sphere}
\end{equation}
where $T_{P\ell;np}$ is the, diagonal with respect to $L$, element of the $T$~matrix.
Most results in the following are given in terms of those cross sections.

\section{\label{results}Results and Discussion}
Recent advances in low-loss metaphotonic platforms have concentrated
on all-dielectric structures supporting Mie-type
resonances~\cite{koshelevDielectricResonantMetaphotonics2021}.
Dynamic control through temporal modulation of such structures
has further enabled the engineering of inter-resonance
coupling~\cite{sadafiDynamicControlLight2023}.
Efficient inelastic scattering in resonant systems
requires substantial spatial overlap between the electromagnetic fields
at the incoming and outgoing frequencies. This requirement can be met
by using tightly confined, symmetry-matched modes,
especially when both frequencies correspond
to resonant states of the structure~\cite{%
pantazopoulosLayeredOptomagnonicStructures2019a,%
pantazopoulosHighefficiencyTripleresonantInelastic2019,%
almpanisSphericalOptomagnonicMicroresonators2020}.

Although the scattering $T$~matrix formalism has previously been applied
to particles with periodic time-modulation,
a simplified Born approximation framework can offer
valuable physical insight and guide the optimization
of inelastic scattering processes. Here, we apply this approach
to two geometries: a dielectric sphere,
for which the Born approximation admits an analytical solution,
and a dielectric cylinder, which requires a numerical treatment
based on \eqrefdio{eq:born_tmatrix_nodisp}. For the cylindrical case,
we specifically investigate inelastic scattering mediated
by a so-called \emph{supercavity} mode~\cite{rybinHighSupercavityModes2017}.

\subsection{\label{subsection:sphere_results}Dielectric sphere}
Light scattering from dielectric spheres with harmonically varying permittivity
or radius, at frequencies in the vicinity of a Mie resonance,
has been analyzed in previous studies~\cite{stefanouLightScatteringSpherical2021,%
panagiotidisInelasticLightScattering2022a,%
panagiotidisOpticalTransitionsNonreciprocity2023a}.
The presence of a resonance can significantly enhance the generated harmonics
and induce asymmetries between the Stokes and anti-Stokes components.
Such asymmetries enable an exchange of energy between the optical field
and the external mechanism responsible for driving the periodic permittivity variation.

\begin{figure}
\resizebox{\columnwidth}{!}{\includegraphics{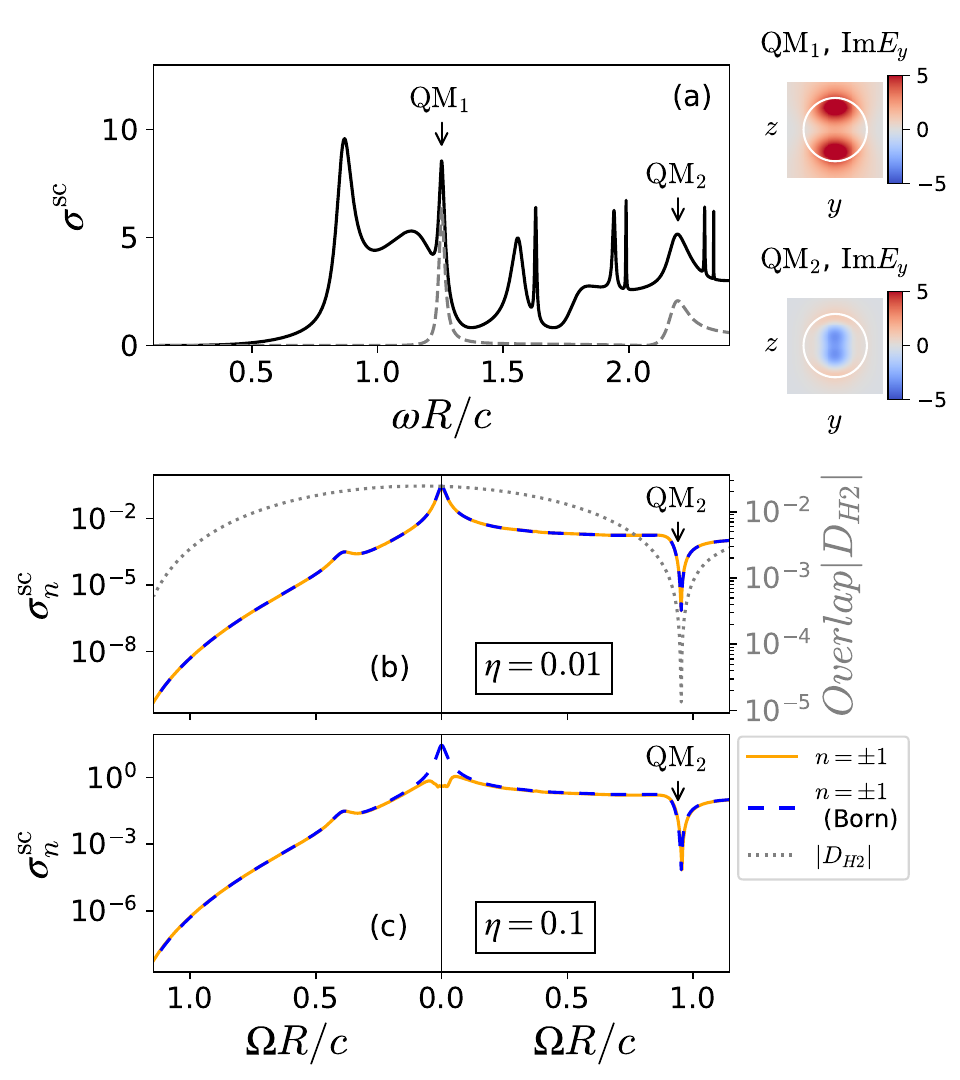}}
\caption{\label{fig:sphere} Light scattering by a sphere of radius $R$ with static
permittivity $\varepsilon = 12$
and permeability $\mu = 1$.
Spherical waves up to $\ell_{\rm{max}} = 5$ have been retained.
(a): Static scattering cross section of a plane wave, incident along the
$x$ direction, calculated by the static \ebcm\ (solid line),
and its $P\ell = H2$ component (dashed line).
We can see two resonances, $\text{QM}_{1}$ and $\text{QM}_{2}$,
corresponding to $P\ell = H2$, in the displayed spectral region.
In the margin, we show the imaginary part of the $y$ component
of the electric field on the $y$-$z$ section passing through the center of the sphere,
for scattering at these resonances.
(b): First inelastic ($n = \pm 1$) scattering cross section components
for incoming frequency on resonance $\text{QM}_{1}$,
as we vary the modulation frequency.
The right (left) parts correspond to $n = -1$ ($n = 1$).
The full (dashed) lines show the dynamic \ebcm\ (Born approximation) results.
A sharp dip for $n = -1$ is clearly visible.
The dotted line depicts the absolute value of
the radial overlap integral $|D_{H2}|$ from \eqrefdio{eq:overlap_sphere_H}.
(c): Same as (b), but with higher modulation amplitude.}
\end{figure}

Figure~\ref{fig:sphere}a shows the scattering cross section
of a linearly polarized plane wave incident
on a static dielectric sphere of radius $R$,
relative permittivity $\varepsilon = 12$, and permeability $\mu = 1$.
The spectral response exhibits prominent contributions from Mie resonances,
each characterized by its angular momentum index $\ell$
and polarization (electric or magnetic). As energy increases,
resonances associated with higher $\ell$ values emerge,
along with higher-order radial modes for a given $\ell$.
The resulting spectral overlap leads to a rich optical spectrum.
As discussed above, the spherical symmetry of the system
prevents coupling between modes of different $\ell$
under a uniform time-periodic variation of the permittivity.
Therefore, we restrict our analysis to the first two resonances
for $\ell = 2$ , denoted $\text{QM}_{1}$ and $\text{QM}_{2}$
in \figref{fig:sphere}a. The individual contribution of
only the quadrupolar ($\ell = 2$) modes of magnetic type ($P = H$)
is shown in \figref{fig:sphere}a by the dashed line.
The imaginary part of the $E_{y}$ component for $\text{QM}_{1}$
and $\text{QM}_{2}$ is also depicted on the right-hand side of 
\figref{fig:sphere} as it provides the dominant contribution
to the electric field for both modes.

Next, we introduce a uniform harmonic variation of the permittivity of the sphere of the form
\begin{equation}
\varepsilon(t) = \varepsilon_{0}\varepsilon[1 + \eta\cos(\Omega t)]
\label{eq:modulation}
\end{equation}
and examine the first,
Stokes and anti-Stokes, inelastic components $n = \pm 1$
(with frequencies $\omega \mp \Omega$) of the scattering cross section.
The scattering cross section for incoming frequency
on resonance $\text{QM}_{1}$ is shown in \figrefs{fig:sphere}(b, c),
as we vary the modulation frequency $\Omega$,
for two modulation amplitudes $\eta$.
Both the full dynamic calculation (full lines) and the
Born approximation (dashed lines) results are displayed.
The Born approximation works 
very well for moderate modulation amplitudes ($\eta = 0.01$),
while for larger amplitudes ($\eta = 0.1$) only near the
adiabatic limit---that is, for small $\Omega$---we observe a significant divergence.

The incoming frequency is chosen on resonance~$\text{QM}_{1}$ to maximize
the field corresponding to the input frequency in the overlap integral,
thereby enhancing the inelastic scattering.
By applying the same logic to the field corresponding
to the output frequency, we expect stronger inelastic effects when
the outgoing frequency lies within the same resonance halfwidth, or
corresponds to another resonance with the same angular symmetry ($P, \ell$)
but a different radial order. This likely explains the divergence observed
in the adiabatic limit: a resonance overlaps most strongly with itself,
lowering the threshold at which higher-order phenomena can no longer be neglected.

The most intense inelastic phenomena are indeed observed
in the adiabatic limit and when the modulation frequency lies
within the Mie resonance halfwidth. However, when the modulation frequency
approaches the frequency difference
$\omega_{\text{QM}_{2}} - \omega_{\text{QM}_{1}}$, we observe a pronounced
reduction in the corresponding inelastic scattering cross section,
contrary to the intuitive expectation of enhancement. 
This dip can be understood within the Born approximation and arises
from a node in the radial overlap integral $D_{H2}$
that enters \eqrefdio{eq:born_tmatrix_sphere_nodisp},
which is shown in \figref{fig:sphere}b by the dotted line.
Such a node is expected from the field profiles
shown in \figref{fig:sphere} for incident plane waves.
For the $\text{QM}_{1}$ resonance the sign
of $\imag E_{y}$---the dominant electric-field component,
as noted above---remains constant,
while for $\text{QM}_{2}$ it changes sign near the sphere boundary. 
Consequently, their overlap integral largely cancels.
This overlap integral should approximate $D_{H2}$,
since on resonance the scattering is dominated by spherical waves with $P\ell = H2$.

It is noteworthy that the overlap integral of the two modes
with different radial order is found to vanish for modulation frequencies
slightly offset from $\Omega = \omega_{\text{QM}_{2}} - \omega_{\text{QM}_{1}}$.
This is most likely due to the fact that the temporal variation shifts the
resonances of the time-modulated system with respect to those of
the static system. This shift is best explained through the framework
of quasinormal mode theory~\cite{%
kristensenModelingElectromagneticResonators2020,%
sauvanNormalizationOrthogonalityCompleteness2022}
or resonant state expansion~\cite{%
bothResonantStatesTheir2022}. Both approaches focus on the \emph{complex}
eigenfrequencies and the corresponding
eigenfields of Maxwell's equations in an open system,
and allow for the calculation of resonance frequencies
of a perturbed system from the resonances of a background system ---
the former through coupled mode theories~\cite{%
kristensenModelingElectromagneticResonators2020,taoCouplingTheoryQuasinormal2020,%
zhangQuasinormalCoupledMode2020} and the latter immediately,
through a Mittag-Leffler expansion of the background Green's function.
In both cases, the perturbation causes the resonances to move
in the complex plane, which is sufficient to explain the effect
observed here. Although both theories have been developed for static systems,
a rather recent work~\cite{valeroResonantStatesStructured2025}
generalizes the resonant state expansion to time-modulated Floquet systems.
As described there, Floquet's theorem implies that each static resonance
generates a series of \emph{replicas}, shifted
by integer multiples of the modulation frequency $\Omega$.  The integer $n$
labels these Floquet harmonics, with $n=0$ corresponding
to the original static resonance. In the absence of modulation,
replicas with different $n$ are uncoupled. Modulation, however,
induces interactions between them. In resonance-to-resonance transitions,
the real frequency of one static resonance coincides
with the $n = \pm1$ replica of another, facilitating their interaction
and likely causing the observed shift.

The suppression of harmonic modes is also observed for other pairs of resonances
sharing the same angular momentum and polarization,
but different radial order. However, in dielectric spheres,
the permittivity modulation frequencies required to induce
these effects are relatively high, being comparable to the frequencies
of the Mie resonances themselves. This imposes significant practical limitations,
as achieving permittivity variations
on the order of several terahertz remains challenging.

\subsection{\label{subsection:cylinder}Dielectric cylinder}
To investigate analogous phenomena at modulation frequencies substantially lower
than the frequency of the electromagnetic wave, and to further explore the potential
of resonance-to-resonance transitions, we turn our attention
to a cylindrical scatterer. In this geometry, an additional degree of freedom,
the aspect ratio, enables closer spacing of resonances
and allows for controlled modification of their modal profiles.

\begin{figure}
\resizebox{\columnwidth}{!}{\includegraphics{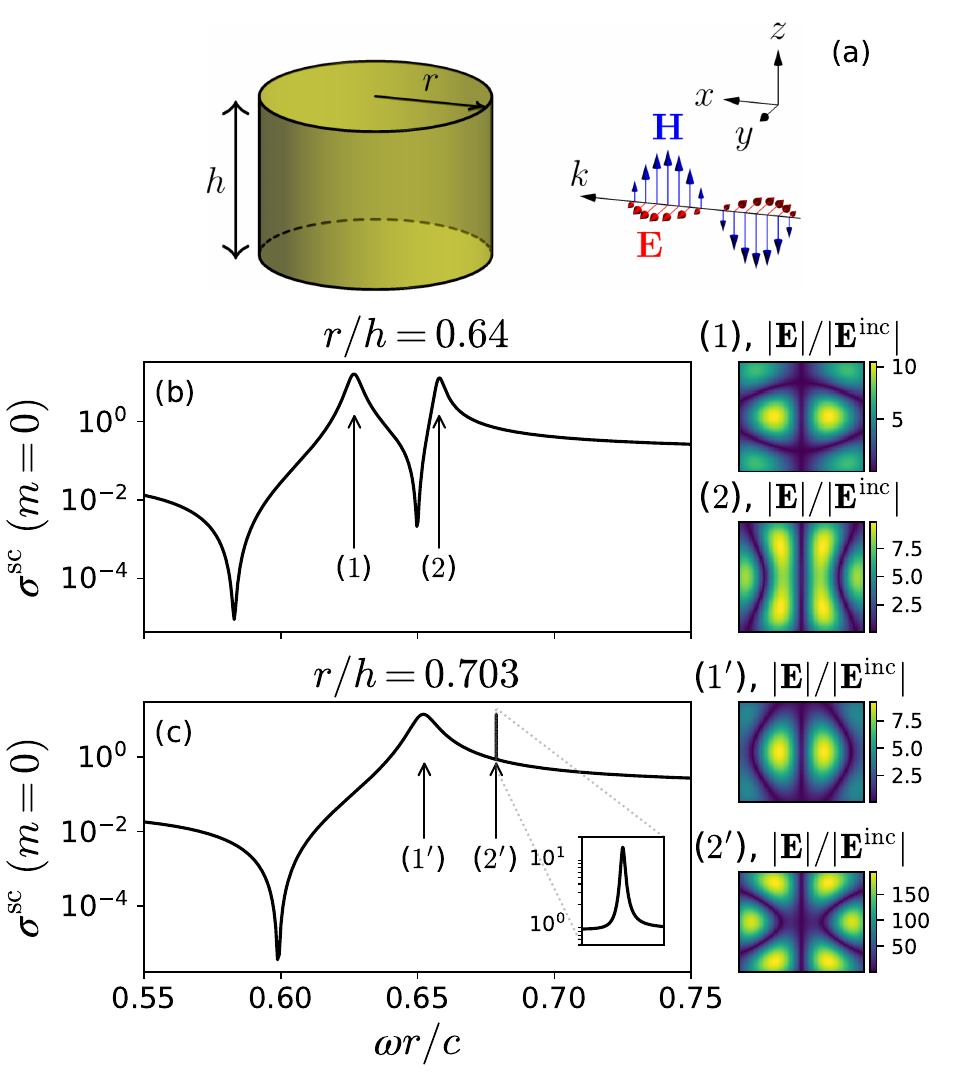}}
\caption{\label{fig:static_m=0_s_pol}
(a): Cylindrical scatterer with $\varepsilon = 80$ and $\mu = 1$,
of radius $r$ and height $h$, illuminated by an s-polarized plane wave
perpendicular to the cylinder axis. (b) and (c):
Static scattering cross sections ($m = 0$ component)
for two different cylinder aspect ratios, plotted
as a function of the dimensionless frequency $\omega r/c$.
The results are obtained using the static \ebcm\ with angular-momentum
cutoffs $\ell_{\rm{max}} = 10$
and $\ell_{\rm{cut}} = 12$~\cite{stefanouLightScatteringPeriodically2023a}.
The four resonances indicated by arrows occur at:
$\omega r/c = 0\decsign6269$ ($1$), $\omega r/c = 0\decsign6579$ ($2$),
$\omega r/c = 0\decsign6525$ ($1'$), and $\omega r/c = 0\decsign67875$ ($2'$).
Resonance ($2'$) corresponds to the supercavity mode.
The normalized electric-field distribution inside the cylinder
in the $y$-$z$ plane, perpendicular to the incident direction $x$,
is shown in the margin for each resonance.}
\end{figure}

We consider a cylinder of radius $r$, height $h$,
and permittivity $\varepsilon = 80$, as described
in~\citewithref{rybinHighSupercavityModes2017}, and examine the scattering
of an $s$-polarized plane wave, incident perpendicular
to the symmetry axis of the cylinder, as shown in \figref{fig:static_m=0_s_pol}a.
The component $m = 0$ of the scattering cross section
is depicted in \figrefs{fig:static_m=0_s_pol}(b, c),
for two different aspect ratios,
$r/h = 0\decsign 64$ and $r/h = 0\decsign 703$, respectively.
The two resonances, labeled (1) and (2) in \figref{fig:static_m=0_s_pol}b
for $r/h = 0\decsign 64$, move closer together as the aspect ratio increases
and, at $r/h = 0\decsign 703$, the supercavity mode,
denoted as resonance~($2'$), is formed,
as demonstrated in~\citewithref{rybinHighSupercavityModes2017}.

\begin{figure}
\resizebox{\columnwidth}{!}{\includegraphics{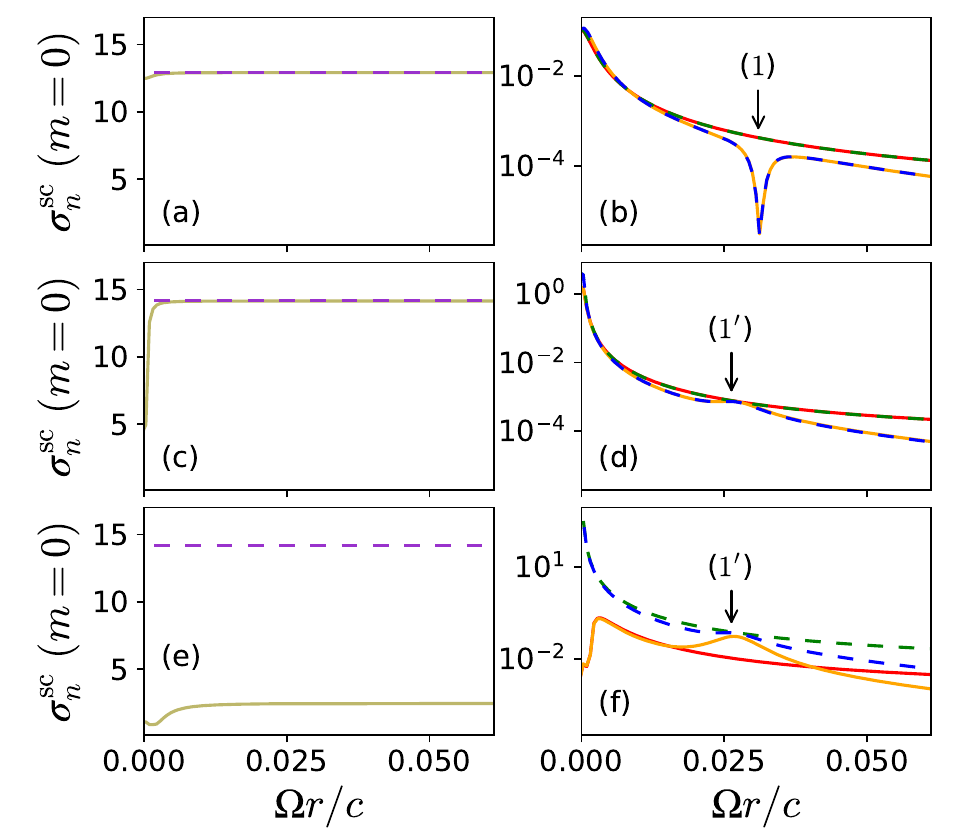}}
\caption{\label{fig:scan_mod_freq_born}Elastic ($n = 0$)
and first inelastic ($n = \pm 1$)
$m = 0$ components of the scattering cross section for the two different aspect
ratios shown in \figref{fig:static_m=0_s_pol}.
We consider an incoming plane wave with frequency
corresponding to resonances ($2$) and ($2'$), respectively,
and vary the dimensionless modulation frequency $\Omega r/c$.
(a): $r/h = 0\decsign64$, $n = 0$, $\eta = 0\decsign001$.
(b): $r/h = 0\decsign64$, $n = \pm1$, $\eta = 0\decsign001$.
(c): $r/h = 0\decsign703$, $n = 0$, $\eta = 0\decsign001$.
(d): $r/h = 0\decsign703$, $n = \pm1$, $\eta = 0\decsign001$.
(e): $r/h = 0\decsign703$, $n = 0$, $\eta = 0\decsign01$.
(f): $r/h = 0\decsign703$, $n = \pm1$, $\eta = 0\decsign01$.
The full (dashed) lines show the dynamic \ebcm\ (Born approximation) results.
For the figures on the right, the red and green curves show $n = -1$,
while the orange and blue curves show $n = +1$.
It can be seen that, for $r/h = 0\decsign703$,
a peak appears when we have resonance-to-resonance transition;
that is, when the modulation frequency is such that the $n = 1$ component
corresponds to resonance ($1'$). On the contrary,
for $r/h = 0\decsign64$
a sharp dip appears when the $n = 1$ component
corresponds to resonance ($1$).}
\end{figure}

We next introduce, as before, a time-variation
of the permittivity of the cylinder in the form of \eqrefdio{eq:modulation},
and study the scattering of incoming  
plane waves with frequencies on resonance ($2$) or ($2'$).
The Stokes and anti-Stokes components $n = \pm 1$
of the scattering cross section are depicted in \figref{fig:scan_mod_freq_born},
along with the elastic ($n = 0$) component, which serves as a further test
of the validity of Born approximation.

In the adiabatic limit, the system is anticipated to exhibit
the most pronounced inelastic phenomena. Furthermore,
the transition between selected resonances,
induced by a harmonic permittivity variation,
is fundamentally governed by the overlap integral
between the respective mode profiles.
This is indeed confirmed by the results shown in \figref{fig:scan_mod_freq_born}.
For an aspect ratio $r/h = 0\decsign64$, the $n = 1$ (Stokes) component
of the scattering cross section (\figref{fig:scan_mod_freq_born}b)
exhibits a sharp dip at $\Omega \approx \omega_2 - \omega_1$.
In contrast, for $r/h = 0 \decsign 703$,
when the modulation frequency matches the frequency difference
between resonances ($2'$) and ($1'$) (see \figref{fig:scan_mod_freq_born}d),
the destructive interference is suppressed,
leading to a broad peak of moderate intensity.
Since the Born approximation accurately reproduces the exact results
for a modulation amplitude $\eta = 0\decsign001$, this behavior should be attributed
to a modification of the modal overlap. This reconfiguration arises
from the strong interaction between resonances.

For $r/h = 0\decsign703$ and $\eta = 0\decsign01$,
the Born approximation is no longer reliable as evidenced
from \figrefs{fig:scan_mod_freq_born}(e, f).
However, the peak just above $\Omega r/c =0.025$ still survives and almost surpasses
the inelastic cross section for $\Omega \rightarrow 0$.
Transitions between resonances are difficult to achieve in Floquet systems
with single symmetric particles, such as spheres or cylinders.
However, stronger modulation or alternative geometries, for instance collections of particles,
can produce pronounced resonance-to-resonance transitions,
as predicted for layers of spherical
particles~\cite{panagiotidisOpticalTransitionsNonreciprocity2023a}.

\begin{figure}
\resizebox{\columnwidth}{!}{\includegraphics{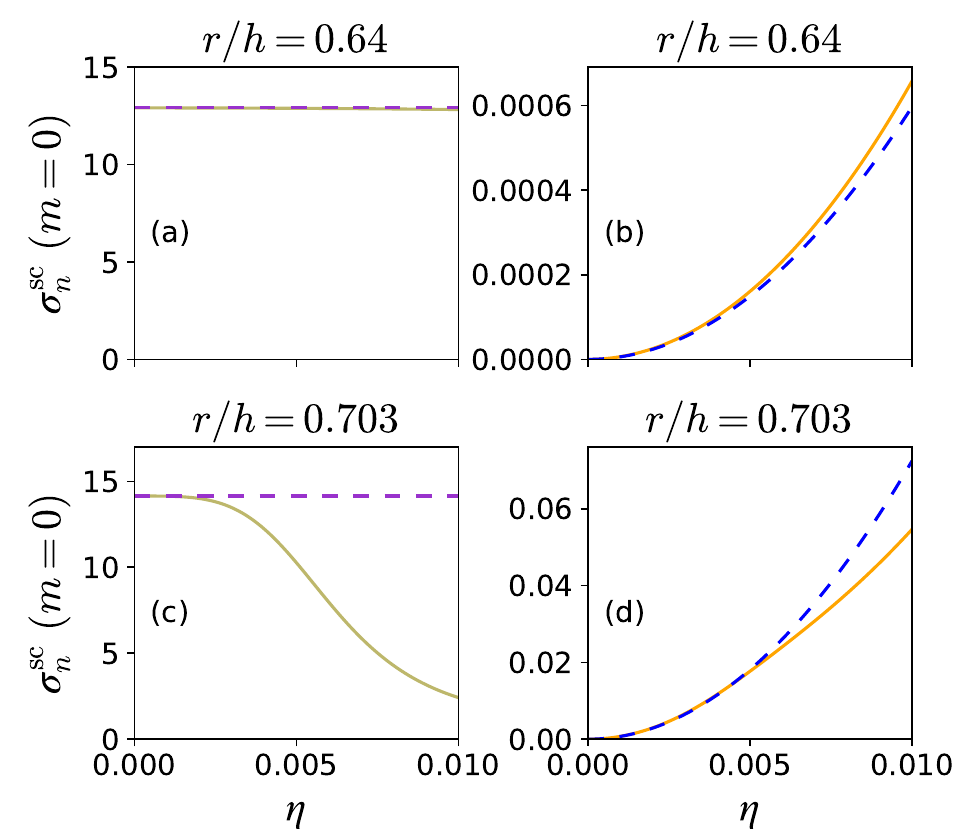}}
\caption{\label{fig:varampl}Elastic ($n = 0$, left-hand diagrams)
and inelastic ($n =1$, right-hand diagrams)
components of the scattering cross section
as a function of the modulation amplitude $\eta$ for the transitions
($2$) $\rightarrow$ ($1$) and ($2'$) $\rightarrow$ ($1'$),
manifested in the right-hand diagrams of \figref{fig:scan_mod_freq_born}.
The full (dashed) lines show the dynamic \ebcm\ (Born approximation) results.
For both aspect ratios, the Born approximation breaks down at sufficiently large $\eta$.}
\end{figure}

As another test of the limits of the Born approximation,
we examine the resonance-to-resonance transition
for $n = 1$, corresponding to the points labeled ($1$) and ($1'$)
in the right-hand diagrams of \figref{fig:scan_mod_freq_born},
as the modulation amplitude is increased.
In \figref{fig:varampl} we show the variation of the elastic ($n = 0$)
and inelastic $n = 1$ components of the scattering cross section
as we increase the modulation amplitude $\eta$.
We see that the Born approximation is valid at
small modulation amplitudes ($\eta \sim 10^{-3}$)
but higher-order effects quickly become important as $\eta$ increases,
causing the approximation to fail.

\section{\label{conclusions}Conclusions}
In summary, within a perturbative framework, we formulated
a first-order Born approximation for light scattering
by particles with a permittivity that varies periodically in time.
By treating the temporal variation as a weak perturbation
to an otherwise static particle, as appropriate in practical implementations,
we obtained compact expressions for the $T$~matrix that clarify
how inelastic scattering amplitudes are governed by overlap integrals
between the modes at the input and output frequencies.
This insight offers a clear physical interpretation
of frequency-conversion processes and directly connects the strength
of inelastic channels to the spatial structure and mutual overlap
of the underlying static modes. 

Using a dielectric sphere as a benchmark system, we demonstrated that
the Born approximation quantitatively reproduces dynamic \ebcm\ results
for modest modulation amplitudes and accurately captures both enhancement
and suppression of inelastic scattering
for resonance-to-resonance optical transitions,
including sharp dips due to near-orthogonality of radial modes.
Extending the analysis to a high-permittivity cylindrical resonator,
we showed that geometric tuning of the aspect ratio brings resonances
close enough to hybridize, altering their modal overlap
and transforming dips in inelastic scattering into prominent peaks.

These results highlight how the spatial structure and mutual overlap
of static modes directly govern the strength of inelastic channels
and demonstrate that simple geometric or material tuning
can selectively suppress or enhance frequency-conversion processes
in time-modulated photonic resonators. Moreover, the approach
can naturally account for nonuniform refractive index-variations
and can be readily incorporated into existing static $T$~matrix formulations
to describe scattering from time-modulated scatterers. Overall,
this work confirms the predictive power of the perturbative approach,
clarifies the central role of mode structure in time-dependent scattering,
and provides practical guidance for designing dynamically modulated resonators
with tailored frequency-conversion properties.

Future extensions may incorporate higher-order perturbative corrections,
explore nonuniform or anisotropic time-variations,
and integrate the present formalism into multiple-scattering frameworks
to describe complex time-modulated photonic architectures.

\appendix

\section{\label{appendix:overlap}Calculation of the volume integral}
Using the identity
\begin{equation}
	\label{eq:vector_green_identity}
	\begin{split}
		 & \int_{V} \left[\vecdio{a} \cdot
         (\vecnabla \times \vecdio{\nabla} \times \vecdio{b})
		- (\vecnabla \times \vecnabla \times \vecdio{a})
        \cdot \vecdio{b}\right] \,  \mathrm{d}^{3}r \\
		 & = -\int_{\partial V}  \left\{\left[\unitvec{n} \times
         (\vecnabla \times \vecdio{a})\right] \cdot \vecdio{b}
		+ (\unitvec{n} \times \vecdio{a})\cdot
        (\vecnabla \times \vecdio{b})\right\} \, \mathrm{d}^{2}r
	\end{split}
\end{equation}
and the following properties of vector spherical waves
\begin{subequations}
	\begin{align}
		 & L = p\ell m \Leftrightarrow
         \polconj{L} = 1 - p, \ell, m\eqcomma       \\
		 & \vecnabla \times \vecdio{J}_{L}(\alpha, \vecdio{r})
		= (-1)^{p} i\alpha\vecdio{J}_{\polconj{L}}(\alpha, \vecdio{r})\eqcomma       \\
		 & \vecnabla \times \vecnabla \times \vecdio{J}_{L}(\alpha, \vecdio{r})
		= \alpha^{2} \vecdio{J}_{L}(\alpha, \vecdio{r})\eqcomma
	\end{align}
\end{subequations}
we obtain
\begin{equation}
	\begin{split}
		\int_{V_{\text{in}}}& \vecdio{J}_{L}(\alpha, \vecdio{r}) \cdot
        \vecdio{J}_{L'}(\beta, \vecdio{r}) \, \mathrm{d}^{3}r                       \\
		=& \frac{i}{\beta^{2} - \alpha^{2}}\int_{\partial V_{\text{in}}} \unitvec{n}\cdot
		\Big\{(-1)^{p}\alpha\vecdio{J}_{L'}(\beta, \vecdio{r}) \times 
        \vecdio{J}_{\polconj{L}}(\alpha, \vecdio{r})                  \\
		& - (-1)^{p'}\beta\vecdio{J}_{L}(\alpha, \vecdio{r}) \times
        \vecdio{J}_{\polconj{L'}}(\beta, \vecdio{r})
		\Big\} \, \mathrm{d}^{2}r \eqendpoint
	\end{split}
\end{equation}
Introducing the $J^{+}$ matrix used in the
\ebcm~\cite{stefanouLightScatteringPeriodically2023a}
\begin{equation}
	J^{+}_{LL'}(\alpha, \beta) = \alpha\beta\int_{\partial V_{\text{in}}} \unitvec{n}\cdot
	\left[\vecdio{J}_{L'}(\beta, \vecdio{r}) \times \angconj{\vecdio{J}}_{L}(\alpha, \vecdio{r})
    \right] \, \mathrm{d}^{2}r\eqcomma
\end{equation}
we can rewrite the integral as
\begin{equation}
\begin{split}
	&\int_{V_{\text{in}}} \vecdio{J}_{L}(\alpha, \vecdio{r}) \cdot
    \vecdio{J}_{L'}(\beta, \vecdio{r}) \, \mathrm{d}^{3}r\\
		&= \frac{i}{\beta^{2} - \alpha^{2}} \left\{
        \frac{(-1)^{m}}{\beta}J^{+}_{\angconj{\polconj{L}}L'}(\alpha, \beta)
		- \frac{(-1)^{m'}}{\alpha}J^{+}_{\angconj{\polconj{L'}}L}(\beta, \alpha)
        \right\}\eqendpoint
\end{split}
\end{equation}
Routines to calculate the $J^{+}$ matrices are already in use for the \ebcm\ 
and can be applied straightforwardly.

\section{\label{appendix:sphere}Born approximation for a sphere - Coefficients and integrals}
Setting
\begin{subequations}
\begin{align}
    A &= h^{+}_{\ell}(q_{\text{h}}R)\frac{\partial}{\partial r}
    \left[rj_{\ell}(q_{\text{h}}r)\right]|_{r = R}\eqcomma\\
    B &= j_{\ell}(q_{\text{h}}R)\frac{\partial}{\partial r}
    \left[rh^{+}_{\ell}(q_{\text{h}}r)\right]|_{r = R}\eqcomma\\
    C &= h^{+}_{\ell}(q_{\text{h}}R)\frac{\partial}{\partial r}
    \left[rj_{\ell}(qr)\right]|_{r = R}\eqcomma\\
    D &= j_{\ell}(qR)\frac{\partial}{\partial r}
    \left[rh^{+}_{\ell}(q_{\text{h}}r)\right]|_{r = R}\eqcomma
\end{align}
\end{subequations}
we obtain
\begin{equation}
	\alpha_{E\ell}(k)
	= \frac{qR}{q_{\text{h}}R} \varepsilon_{\text{h}}\frac{A - B}{C\varepsilon_{\text{h}} - D\varepsilon}\eqcomma
\end{equation}
and
\begin{equation}
	\alpha_{H\ell}(k) = \mu
    \frac{A - B}{C\mu_{\text{h}} - D\mu}\eqendpoint
\end{equation}
We also need to calculate the integral
$\int_{V_{\text{in}}}\angconj{\vecdio{J}}_{L'}(q_{n'}, \vecdio{r}) \cdot
\vecdio{J}_{L}(q_{n}, \vecdio{r}) \, \mathrm{d}^{3}r$.
The angular integration gives $\delta_{LL'}$, while the radial integration
can be performed using the identity \cite{gradshteynTableIntegralsSeries2007}
\begin{equation}
	\int_{0}^{1} x^{2}j_{\ell}(\alpha x)j_{\ell}(\beta x) \, \mathrm{d}x =
	\frac{\beta j_{\ell - 1}(\beta)j_{\ell}(\alpha) - \alpha j_{\ell - 1}(\alpha)j_{\ell}(\beta)}
    {\alpha^{2} - \beta^{2}}  \eqendpoint
\end{equation}
Finally we obtain \eqrefdio{eq:born_tmatrix_sphere_nodisp}, where
\begin{equation}
\label{eq:overlap_sphere_E}
	D_{E\ell}(\alpha, \beta) = \frac{\alpha j_{\ell - 1}(\beta)j_{\ell}(\alpha) - \beta j_{\ell - 1}(\alpha)j_{\ell}(\beta)}
    {\alpha^{2} - \beta^{2}} -\ell\frac{j_{\ell}(\alpha)}{\alpha}\frac{j_{\ell}(\beta)}{\beta}\eqcomma
\end{equation}
and
\begin{equation}
\label{eq:overlap_sphere_H}
	D_{H\ell}(\alpha, \beta) = \frac{\beta j_{\ell - 1}(\beta)j_{\ell}(\alpha) - \alpha j_{\ell - 1}(\alpha)j_{\ell}(\beta)}
    {\alpha^{2} - \beta^{2}}\eqendpoint
\end{equation}
Therefore, every quantity in \eqrefdio{eq:born_tmatrix_sphere_nodisp}
is accessible in a closed form.

\bibliography{bibliography.bib}

\end{document}